\documentclass[pop,aip,preprint]{revtex4}
\usepackage{epsfig}
\usepackage{graphicx}
\usepackage{graphics}
\usepackage{hyperref}

\begin{document}

\title{Quantum Trivelpiece-Gould waves in a magnetized dense plasma}

\author{H. Ter\c{c}as$^{1}$}
\email{htercas@cfif.ist.utl.pt}
\author{J. T. Mendon\c{c}a$^{1,2}$}
\email{titomend@ist.utl.pt}
\affiliation{$^1$CFIF and $^2$IPFN, Instituto Superior T\'{e}cnico, Av. Rovisco Pais 1, 1049-001 Lisboa, Portugal}

\author{P. K. Shukla$^3$}
\affiliation{$^3$Institut f\"{u}r Theoretische Physik IV, Ruhr-Universit\"at Bochum, D-44780 Bochum, Germany}

\begin{abstract}
The dispersion relation for the electrostatic waves below the electron plasma frequency 
in a dense quantum plasma is derived by using the magnetohydrodynamic model. It is shown 
that in the classical case the dispersion relation reduces to the expression obtained for 
the well-known Trivelpiece-Gould (TG) modes.  Attention is also devoted to the case of solitary 
waves associated with the nonlinear TG modes.  
\end{abstract}

\maketitle

\section{Introduction}
In recent years, special attention has been paid to the quantum effects in plasmas. The advent of 
miniaturization techniques \cite{markowich}, the progress on the understanding of the properties of 
dense astrophysical matters \cite{melrose1, melrose2, marklund}, experiments on extreme physics 
like laser plasmas \cite{kremp} and the recent progress in producing ultracold plasmas in terms 
of Rydberg states \cite{li, fletcher}, have envisaged the application of plasma physics where 
the dimensions of the system are frequently comparable to the de Broglie length. Hence, quantum 
mechanical effects, as electron tunneling, nonlocality and interference, are expected to play 
an important role in the dynamics of such dense systems. For that reason, efforts have been made 
to establish models for describing quantum plasmas \cite{haas,manfredi} and to predict new 
interesting phenomena. The latter include quantum beam-plasma instabilities \cite{anderson, haas2, haas3}, 
quantum ion-acoustic waves \cite{haas4}, quantum Alfv\'en waves \cite{shukla1} and quantum surface 
waves \cite{shokri}. The evidence of nonlinear phenomena has also been reported, e.g. quantum corrections 
to the Zakharov equations \cite{garcia} and magnetosonic solitons \cite{marklund2}. The exciting fundamental 
work has also been done in what  concerns the  kinetic theory of plasmas, where generalizations of the 
classical transport equations, based on the Wigner formalism, have been performed, envisaging 
a phase-space description of quantum systems. As a consequence, it is possible to explore purely 
kinetic aspects in quantum plasmas, such as Landau damping and kinetic instabilities.\par

In this paper, we focus on the wave phenomena in bounded quantum plasmas. More specifically, we consider 
the propagation of electrostatic waves, with frequency near the electron plasma frequency, in a magnetized 
plasma column. We will consider linear and nonlinear wave solutions. First, we derive the dispersion relation 
for the quantum version of the Trivelpiece-Gould modes. We show that, in the classical limit, 
$\hbar \rightarrow 0$, the dispersion relation reduces to the expression derived in Ref. \cite{cabral}. 
Second, we will consider the existence of Trivelpiece-Gould solitons in a quantum plasma cylinder 
and discuss the main differences with respect to the classical solutions.

\section{Linear Dispersion Relation}

The motion of the quantum electron fluid, in the presence of a magnetic field $\mathbf{B}=B_0\mathbf{e_z}$, 
is governed by the magnetohydrodynamical equations \cite{haas} 

\begin{eqnarray}
\frac{\partial n}{\partial t}+\mathbf{\nabla}\cdot(n\mathbf{v})&=&0\nonumber\\
\left(\frac{\partial}{\partial t}+\nu_c\right) \mathbf{v}+(\mathbf{v}\cdot\mathbf{\nabla})\mathbf{v}&
=&\frac{e}{m}\left(\mathbf{\nabla} \phi+\mathbf{v}\times\mathbf{B}\right)
-\frac{\mathbf{\nabla} P}{mn}+\frac{\hbar^2}{2m^2}\mathbf{\nabla}
\left(\frac{\mathbf\nabla^2\sqrt{n}}{\sqrt{n}}\right)\label{eq:1}\\
\nabla^2\phi&=&\frac{e}{\epsilon_0}(n-n_0),\nonumber
\end{eqnarray}
where $n_0=n_{e0}$ represents the equilibrium electron number density. Here, we  neglected the motion of 
ions and assumed that the ion number density $n_i = n_0$. 

Let us consider a plasma column of radius $a$. In that case, we can split the Laplace operator into 
the longitudinal and transversal components, $\nabla^2=\nabla_\perp^2+\partial_z^2$, and expand 
the electric potential $\phi$ in terms of Bessel functions

\begin{equation}
\phi(r,\theta,z;t)=\sum_{\ell,m=0}^\infty\phi_{\ell,m}(z;t)J_{\ell,m}(k_{\perp \ell,m} r)e^{im\theta}.
\label{eq:2}
\end{equation}
We assume that the plasma column is surrounded by a conductive wall. 
Because the electric field must vanish outside the column, we expect the wave propagation along the cylinder, 
which corresponds to the mode quantization in the transverse direction, such that

\begin{equation}
k_{\perp \ell,m}=\frac{\alpha_{\ell,m}}{a},
\label{eq:3}
\end{equation}
where $\alpha_{\ell, m}$ stands for the $\ell$th zero of the Bessel function of order $m$, and 
yields the following condition 

\begin{equation}
\left(\nabla_\perp^2+k_{\perp \ell,m}^2\right)J_m(k_{\perp\ell,m}r)e^{im\theta}=0.
\label{eq:4}
\end{equation}
Using Eqs. (\ref{eq:3}) and (\ref{eq:4}), and taking the limit of a very strong magnetic field, such that 
the transverse component of the electron fluid velocity vanishes, $\mathbf{v}_\perp=0$, we can linearize 
Eqs. (\ref{eq:1}) by splitting every relevant physical quantity into its equilibrium and perturbed components, 
$x=x_0+x_1$, and write the following set of one-dimensional equations

\begin{eqnarray}
\frac{\partial n_1}{\partial t}+n_0\frac{\partial v_1}{\partial z}&=&0\nonumber\\
\left(\frac{\partial}{\partial t}+\nu_c\right)v_1&=&\frac{e}{m}\frac{\partial\phi_1}{\partial z}
-\frac{1}{mn_0}\frac{\partial P_1}{\partial z}+\frac{\hbar^2}{4m^2n_0}\frac{\partial^3n_1}{\partial z^3}\\
\left(\frac{\partial^2}{\partial z^2}-k_{\perp(\ell,m)}^2\right)\phi_1&=&\frac{e}{m}n_1.\nonumber
\label{eq:5}
\end{eqnarray}
Here, we have used an approximation for the quantum term (the Bohm potential) 

\begin{equation}
\frac{\hbar^2}{2m^2}\frac{\partial}{\partial z}\left(\frac{\partial_z^2\sqrt{n}}{\sqrt{n}} \right)
\approx\frac{\hbar^2}{4m^2n_0}\frac{\partial^3n_1}{\partial z^3},
\label{eq:6}
\end{equation}
which holds for linear perturbations $n_1\ll n_0$. We now assume that the hydrodynamical pressure of the 
electrons follow an adiabatic law

\begin{equation}
\frac{P}{P_0}=\left(\frac{n}{n_0}\right)^\gamma,
\label{eq:7}
\end{equation}
where $\gamma=(2+d)/d$ represents the adiabatic constant, and $d$ the relevant dimensions. 

Supposing that the waves propagate only along the longitudinal direction, without exchanging any energy along 
the transversal one, we set $\gamma=3$ \cite{manfredi}. Assuming that the electrons are cold enough so the plasma 
can be regarded as a degenerate one-dimensional Fermi gas, in agreement with the strong magnetic field limit, 
we have 

\begin{equation}
P_0=P_F=\frac{1}{3}mn_0v_F^2,
\label{eq:8}
\end{equation}
where $v_F=\hbar/m(3\pi^2n_0)^{1/3}$ represents the Fermi speed. In that case, we can assume a perturbation 
in the pressure of the form

\begin{equation}
\frac{\partial P_1}{\partial z}\approx mv_F^2\frac{\partial n_1}{\partial z}.
\label{eq:9}
\end{equation}
Finally, by putting Eqs. (\ref{eq:5}) and (\ref{eq:9}) together, and performing a Fourier transform,
we obtain the Trivelpiece-Gould dispersion relation for the electrostatic waves propagating along 
the axis of the column

\begin{equation}
\omega^2_{\ell,m}\left(1+i\frac{\nu_c}{\omega_{\ell,m}}\right)
=\omega_p^{2}\frac{k^2}{k^2+k_{\perp(\ell,m)}^2}+v_F^2k^2+\frac{\hbar^2}{4m^2}k^4,
\label{eq:10}
\end{equation}
where $\omega_p$ is the electron plasma frequency. In the classical collisionless regime, $\hbar\rightarrow 0$ 
and $\nu_c\rightarrow 0$, we recover the already reported dispersion relation for electron plasma waves in a 
plasma column of length $L$

\begin{equation}
\omega_{n,\ell,m}^2=\omega_p^2\frac{k_n^2}{k_n^2+k_{\perp(\ell,m)}}+v_{th}^2k_n^2,
\label{eq:11}
\end{equation}
where $k_n=n\pi/L$. Notice that we have replaced the Fermi speed by the classical thermal speed, 
which is the relevant quantity in this classical limit \cite{cabral}. The latter shows that the 
Trivelpiece-Gould waves propagate at frequencies of the order of $\omega_p$.

\section{Trivelpiece-Gould solitons}

In the most general case, we cannot neglect the nonlinear terms in the set of fluid equations. Hence, we 
devote this section to the study of nonlinear Trivelpiece-Gould waves. Following the steps of 
Ref. \cite{cabral}, we insert the development (\ref{eq:2}) in (\ref{eq:1}) and, after multiplying every 
equation by $1/2\pi J_m(k_\perp r)r\exp(-im\theta)$ and integration over the transversal direction, 
we derive the nonlinear set of one-dimensional equations

\begin{eqnarray}
&&\frac{\partial n_{\ell,m}}{\partial t}+n_0\frac{\partial v_{\ell,m}}{\partial z}
+\sum_{\ell',m',\ell'',m''}\beta(\ell,m,\ell',m',\ell'',m'')
\frac{\partial}{\partial z}(n_{\ell',m'}v_{\ell'',m''})=0\nonumber\\
&&\left(\frac{\partial }{\partial t}+\nu_c\right)v_{\ell,m}+\sum_{\ell',m',\ell'',m''}
\beta(\ell,m,\ell',m',\ell'',m'')v_{\ell',m'}\frac{\partial v_{\ell'',m''}}{\partial z}
=\frac{e}{m}\phi_{\ell,m}-\frac{v_F^2}{n_0}\frac{\partial n_{\ell,m}}{\partial z}
+\frac{\hbar^2}{4m^2n_0}\frac{\partial^3 n_{\ell,m}}{\partial z^3}\nonumber\\
&&\left(\frac{\partial^2}{\partial z^2}-k_\perp^2\right)\phi_{\ell,m}=\frac{e}{\epsilon_0}n_{\ell,m},
\label{eq:2.1}
\end{eqnarray}  
where $\beta$ is a nonlinear geometrical factor that couples the different $(m,\ell)$ modes and is given by

\begin{equation}
\beta(\ell,m,\ell',m',\ell'',m'')=\frac{2}{[aJ_{m+1}(\alpha_{\ell,m})]^2}\int_0^aJ_{m'}
(k_{\ell',m'}r)J_{m''}(k_{\ell'',m''}r)J_{m}(k_{\ell,m}r)dr
\label{eq:2.2}
\end{equation}

Without lost of generality, we assume, for the sake of simplicity, that only the fundamental 
mode $(m=0,\ell=1)$ of the cylindrical wave guide is excited, yielding $\beta\approx 0.72$ 
and the one-dimensional fluid equations come

\begin{eqnarray}
&&\frac{\partial n}{\partial t}+n_0\frac{\partial v}{\partial z}
+\beta \frac{\partial }{\partial z}(n-n_0)v=0\nonumber\\
&&\left(\frac{\partial}{\partial t}+\nu_c\right)v+\beta v\frac{\partial v}{\partial z}
=\frac{e}{m}\frac{\partial \phi}{\partial z}-\frac{v_F^2}{n_0}\frac{\partial n}{\partial z}
+\frac{\hbar^2}{4m^2n_0}\frac{\partial^3 n}{\partial z^3}\nonumber\\
&&\left(\frac{\partial^2}{\partial z^2}-k_\perp^2 \right)\phi=\frac{e}{\epsilon_0}\left(n-n_0\right),
\label{eq:2.3}
\end{eqnarray}
where $k_\perp=k_{\perp 1,0}$. We now normalize all the relevant physical quantities

\begin{equation}
n'=\frac{n}{n_0}, \quad v'=v\frac{k_\perp}{\omega_p}, \quad \phi'=\frac{e}{m}\frac{k_\perp^2}{\omega_p^2}\phi,
\label{eq:2.4}
\end{equation}
and the space and time coordinates

\begin{equation}
z'=k_\perp z, \quad t'=t\omega_p,
\label{eq:2.5}
\end{equation}
to write Eqs. (\ref{eq:2.3}) in the form

\begin{eqnarray}
&&\frac{\partial n'}{\partial t'}+\frac{\partial v'}{\partial z'}+\beta\frac{\partial}{\partial z'}(n'v')=0\nonumber\\
&&\left(\frac{\partial}{\partial t'}+\gamma\right)v'+\beta v'\frac{\partial v'}{\partial z'}
=\frac{\partial \phi'}{\partial z'} -\Delta\frac{\partial n'}{\partial z'}+H\frac{\partial^3 n'}
{\partial z'^3}\nonumber\\
&&\left(\frac{\partial ^2}{\partial z'^2}-1 \right)\phi'=n',
\label{eq:2.6}
\end{eqnarray}
where we defined the dimensionless parameters

\begin{equation}
\gamma=\frac{\nu_c}{\omega_p}, \quad \Delta=v_F^2\frac{k_\perp^2}{\omega_p^2}, \quad 
H=\frac{\hbar^2k_\perp^4}{4m\omega_p^2}.
\label{eq:2.7}
\end{equation}

In order to explore the different degrees of nonlinearity in the system, we introduce a small parameter $\epsilon$ 
with respect to which we will perturb the physical quantities. For that task, we define a new set of coordinates

\begin{equation}
\xi=\sqrt{\epsilon}(z'-t'), \quad \tau=\epsilon^{3/2}t'
\label{eq:2.8}
\end{equation}
and performing the expansion

\begin{eqnarray}
x'=\epsilon x_1+\epsilon^2x_2+...,
\label{eq:2.9}
\end{eqnarray}
where $x'$ stands for $n'$, $v'$ and $\phi'$, we can split Eqs. (\ref{eq:2.6}) into different orders of $\epsilon$. 
Neglecting the contribution of terms higher than $\epsilon^2$, we can derive the following equation for the 
nonlinear Trivelpiece-Gould waves 

\begin{equation}
\left(\frac{\partial }{\partial \tau}+\frac{\gamma}{2}\right) n_1+\frac{\Delta}{2}\frac{\partial n_1}{\partial \xi}
+\frac{3}{4}\beta \frac{\partial }{\partial \xi}n_1^2+\frac{1-H}{2}\frac{\partial^3n_1}{\partial \xi^3}=0,
\label{eq:2.10}
\end{equation}
which is formally equivalent to the well-known Kortweg-de Vries equation, which has soliton solutions. 
The different combinations of the parameters map different regimes in a quantum plasma. For the sake of 
illustration, let us consider the case of a ultra-cold collisionless plasma. In this case, the plasma is 
underdense such that $k_BT_F\ll \hbar\omega_p$ and $\nu_c\ll \omega_p$. In that situation, $\gamma=\Delta=0$ 
and the general solution of Eq. (\ref{eq:2.10}) reads \cite{sudam}

\begin{equation} 
n_1(\xi,\tau)=\frac{2u_0}{\beta}\mbox{sech} ^2\left[\frac{1}{2}\sqrt{\frac{2u_0}{1-H}}(\xi-u_0\tau)\right],
\label{eq:2.11}
\end{equation}
where $u_0$ is the normalized propagation velocity of the soliton. The same formal solution was obtained in the classical case $H\rightarrow 0$ by Gradov et al. \cite{gradov} for the nonlinear solitary waves propagating in a magnetized plasma cylinder. 

\section{Conclusion}

In this paper, we have considered linear and nonlinear Trivelpiece-Gould waves in a quantum plasma cylinder. 
We have derived the linear dispersion relation of electrostatic waves propagating below the electron plasma 
frequency $\omega_p$ in a cylindrical strongly magnetized quantum plasma. In the classical limit, we recover 
the well known dispersion relation for the Trivelpiece-Gould waves propagating in a plasma column. 
This work points out for the existence of quantum Trivelpiece-Gould waves. We have also studied the  nonlinear 
wave regime, and established soliton solutions for the magnetized quantum plasma cylinder.

\end{document}